A PROJECT REPORT ON

# 5G CELLULAR-
# AN ENERGY EFFICIENCY PERSPECTIVE

**SUBMITTED BY**
DEVEN PANCHAL

*In Partial Fulfillment
of the Requirements for the Degree of
Master Of Science in the*

*School of Electrical and Computer Engineering,
Georgia Institute of Technology*

*December 2014.*

**GUIDED BY**
PROF. DR. JOHN R. BARRY

# ABSTRACT

While the 5G technology of cellular communications promises great capacity and coverage to access information anywhere and anytime, it is feared to have huge power consumption.

Significant research been has been directed towards solving this problem which exists both on the subscribers' side as well as the operators' side. There have been efforts like predicting traffic, modifying the physical layer etc. towards making the 5G technology more energy efficient.

The aim of this study is to see the technology enablers for 5G from an energy efficiency perspective. Efforts will be made to point out specific areas in 5G cellular where improvements or modifications could make 5G cellular more energy efficient.

# 5G USE-CASES

The 5G technology in wireless communications which is projected to arrive sometime around 2020 will be a big wave. It will not be just another cellular generation. It will be the way the world will be, post-2020.

Let us look at some fascinating applications being attributed to 5G.

**Smart Homes/Smart Buildings/Smart Cities:**

Consider this: A smart energy meter connected to a cloud could help you save electricity bills and sync to your daily schedule in your calendar to heat or cool the house when you leave office for home. Your refrigerator would send you a message as milk level in a carton gets low and detect foods gone bad, and the app on your smartphone could automatically add those items to your shopping list and also show you real time discounts at the local store you purchase groceries at. All without your intervention. How about all the vehicles on the road talking to each other to communicate information like traffic conditions, parking ,accidents, bad road conditions, and compute the best route to the destination also taking into account the amount of gas remaining in your car and also recommend a gas station nearby? Again without you intervening. The thousands of vehicles on the road make sure to see that 2 vehicles don't collide when they are within short distance from each other. On the other hand, they coordinate to streamline traffic flow. How about autonomous vehicles for your personal use?

**Revolutionizing the Medical industry:**

The medical industry will see a huge revolution with applications such as tele surgery becoming commonplace (for which 5G would provide reliable communications), automatic telemetry for all patients, and quick access to a huge repository of high resolution of data like medical images like MRI, CT etc. on a smartphone to doctors.

**Entertainment and life beyond 2020:**

5G holds the promise to change life as a whole. Augmented reality and virtual reality will be combined to deliver an unprecedented gaming experience. 3D-audio and 3D-video would be used to say synchronize an orchestra in the U.S. and a dance troupe in India in real-time to deliver a rocking performance. How about the network moving to the football stadium to provide tens of thousands of users operating UHD devices with the content relevant to them? It is like the internet coming to the user with all the information the user will need, without the user having to go to a place from where he can connect to the internet or get the best speeds.

There are many other use cases which have been envisioned including public safety systems, remote control and monitoring of critical industries and infrastructure without human intervention. There will be even more and revolutionary applications which cannot even be thought of now which 5G will enable.

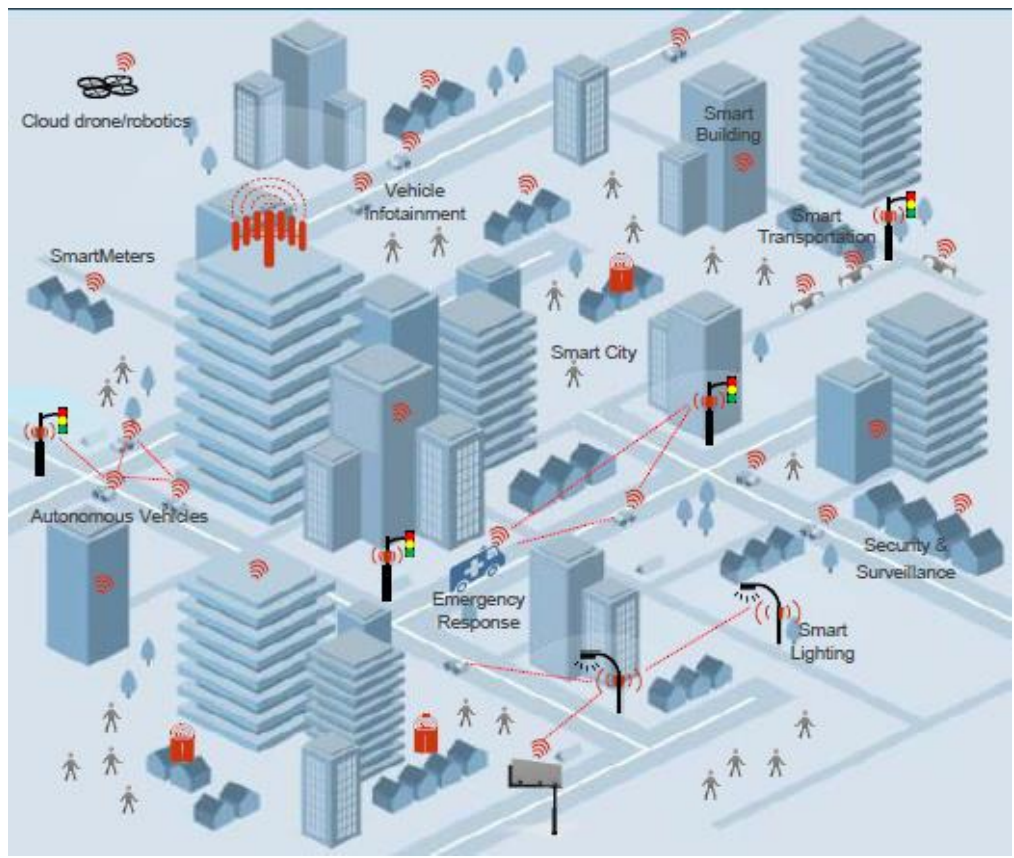

**Fig 1. An illustration showing a city in the 5G-era.** *[Source: Qualcomm].*

In short, it is said that 5G will provide capabilities to connect everyone (networked society) and everything (Internet of Things (IoT)) that can be connected and that too anywhere and at anytime.

It must be borne in mind that the 5G technology will not be just another cellular technology. It will not only span the entire Information and Communications Technology(ICT) ecosystem, but also revolutionize it by creating new business opportunities for and redefining roles of the players in not only ICT, but also in industries that it will touch.

It is clear that 5G technology will be a nation's most important infrastructure, just like bridges and have the highest impact on the nation's economy.

# THE NEED FOR 5G

The spider diagram below shows just some of the reasons why we need 5G.

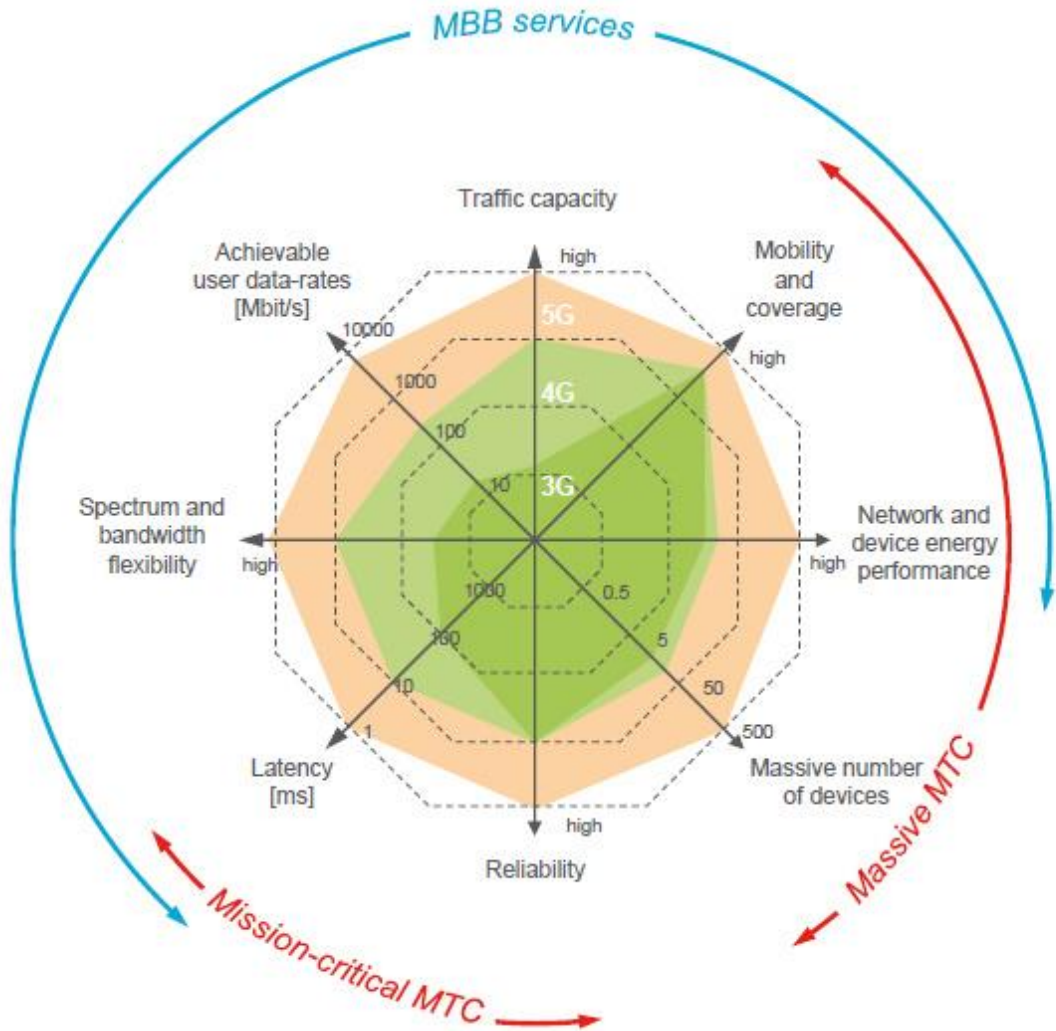

**Fig 2. Spider-diagram showing what 5G would be required to deliver.** *[Source: Ericsson].*

It is clear that the kinds of applications we have discussed ahead cannot be running on the networks we have in place today. By 2020, it has been projected that the data volumes will be more than 1000 times than that of today simply due to the number of data-hungry applications like the ones discussed above. The number of connected devices would grow by more than 100 times with 10 times of battery life required. User data rates will increase by 100 times and will require 5 times lesser latency.

# THE ENERGY PROBLEM FROM THE ICT PERSPECTIVE

Continuously growing carbon emission are aggravating the problem of global warming and climate change- the effects of which the world has witnessed many times. Therefore, energy conservation, is the most pressing issue facing mankind today. Although when compared to other sectors, ICT is responsible for only about 2-2.5% of the global greenhouse gas emissions, this number is growing rapidly as we become a networked society. In 2012, the annual average power consumption of the ICT sector was more than 200GW, 25% of which was due to telecom infrastructure and devices. This has led to the need for energy efficient communication technologies for the future.

Thankfully the industry players have realized this and are seriously working on this and have pledged to reduce carbon emissions. This could be because energy efficiency in ICT does not only offer ecological advantages. It has economical advantages as well. It has great savings for operators since it reduces their Operation Expenditure (OPEX), which can be about 15-35% of the total energy expenditure of the operator. This reduction in OPEX can allow the operator to enable larger infrastructure deployments for capacity upgrades without requiring significant increase in average revenue per user (ARPU) [6]. This directly benefits the user who can now get better services at lesser charges.

There is another reason why we need to think about energy efficiency in 5G. Consider this: According to the 2010 wireless smartphone customer satisfaction survey by JD Power and Associates, the iPhone stood $1^{st}$ in every category except battery life [14]. This means that today's devices are not yet ready to take on the challenge of the applications we discussed earlier. Although efforts have been directed towards developing devices and software to be energy efficient and operate at low power, it is also the job of the network to make sure that it does not drain any device which connects to it. We will subsequently see how different types of devices with very different QoS requirements will come in the way of energy efficiency. Hence low power consumption in the network and in the devices that connect to the network, is an important design target but also a tough engineering challenge as we intend to do more with less.

In the next section, we will go through some of the enabling technologies for 5G and try to understand them from an energy efficiency perspective.

# ENABLING TECHNOLOGIES FOR 5G- AN ENERGY EFFICIENCY PERSPECTIVE

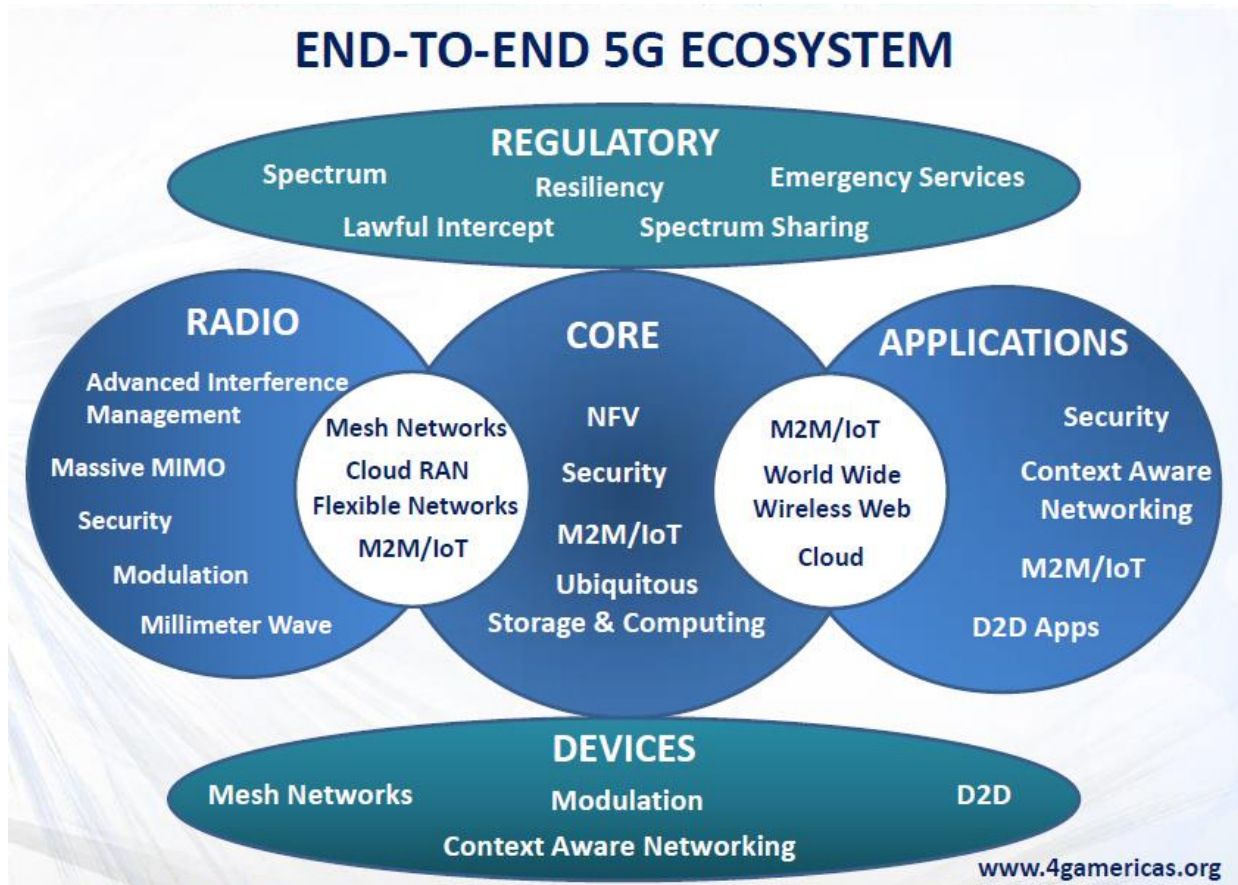

**Fig 3. Enabling Technologies for 5G.** *[Source: [17]].*

The above figure best describes our current understanding of what 5G will be like.

We will start looking at the technologies of the physical layer and move upwards.

**MIMO and Massive-MIMO:**

Although 5G systems will use Massive MIMO for various reasons, MIMO itself inherently provides many energy efficiency benefits. By increasing the network throughput by providing diversity gain or multiplexing gain, it in a way reduces energy consumption. The high spectral efficiency resulting from precoded MIMO translates to high data rates which is equivalent to more sleep time for the node once the data is transmitted. This saving can be huge considering the amount of nodes there will be around considering connected devices due to IoT. But there is another concern too [4]. The total system power consumption is the sum of transmit power and circuit power consumption. Accordingly, the total system power for a MIMO system would have to take into account the DSP blocks too, which add to the circuit power consumption. MIMO is not always energy efficient and for short distances, it may be better to switch to SISO [4].

The MIMO concept can be extended to Multi-user MIMO (MU-MIMO) where multiple users perform local information exchange to achieve distributed transmission and information processing. Although information exchange and CSI knowledge adds to energy consumption, MU-MIMO has been found to be energy efficient compared to SISO uptil certain distances.

Massive MIMO employing way more number of antennas than the number of users can provide energy efficiency in 2 ways-

a. Since it has unused degrees of freedom, these can be used for hardware friendly signal shaping like reduced PAPR signals which can allow use of cheap and power efficient RF amplifiers.
b. Highly selective beamforming at millimeter wave frequencies reduces interference thus reducing wastage of transmitted power.

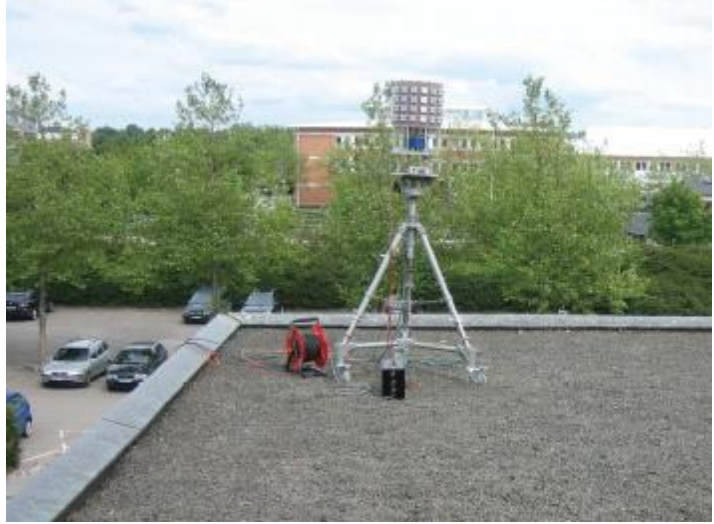

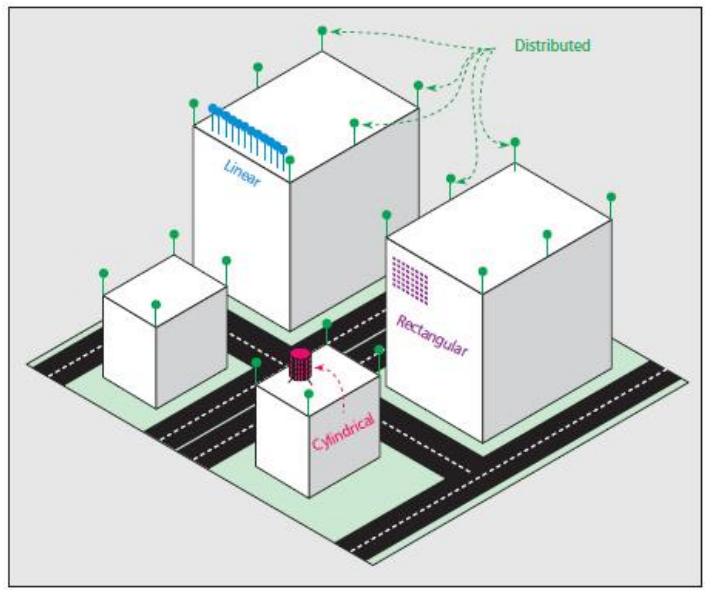

**Fig 4. Possible implementations of Massive-MIMO** *[Source: [13]]*

Massive MIMO has the potential to reduce the transmit power by a factor of 1000.

**C-RAN:**

Distributed architecture for mobile fronthaul and backhaul has evolved from the integrated Baseband Unit (BBU)-Remote Radio Head (RRH) to separate BBU and RRH to BBU-Hoteling and Distributed RRH to BBU Pooling and Distributed RRH. The centralized RAN or cloud RAN (C-RAN) is the next stage in the evolution where the RRH is pushed as close as possible to the user (small-cell) and the BBU is implemented in the cloud. Besides allowing the provisioning of soft services such as Co-ordinated Multipoint (CoMP), multi-RAT virtualization, soft and dynamic cell reconfiguration, its cooling is more energy efficient. The use of C-RAN technology, led to energy savings of 70% in the BS OPEX during 2G and 3G trials in China.

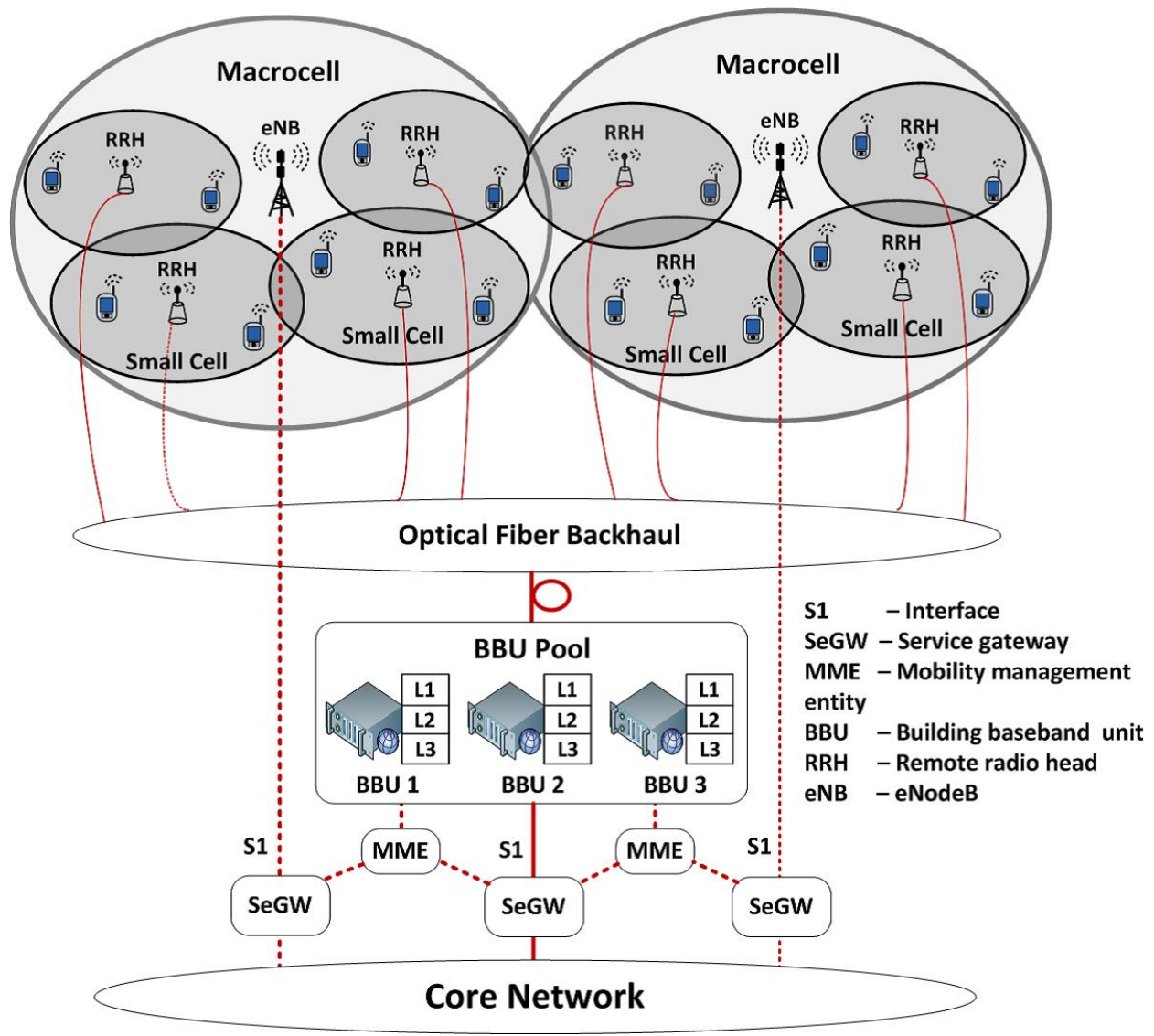

**Fig 5. BBU Pooling- A possible implementation of C-RAN** *[Source: NEC Labs]*

**Waveforms and Modulation:**

Orthogonal Frequency Division Multiplexing (OFDM) has been used in 4G systems and has been the topic of interest worldwide. In [21], an OFDMA Energy efficient scheduler based on Energy efficient link adaptation changing the rate and transmit power and resource i.e. subcarrier allocation is 50% more Energy efficient than a round-robin scheduler. For example, for Machine-to-Machine type communications (M2M) or IoT type traffic, which is sporadic, [10] shows that it is more energy efficient to spread out the transmissions in time.

But for 5G other modulation schemes are also being considered. We can use multicarrier schemes like UFMC for an asynchronous approach that allows open-loop timing control which costs more energy and signaling overhead [10]. UFMC is also more spectrally efficient compared to OFDM. Additionally for the sake of applications with smaller payloads, we are looking at Non-orthogonal multiple access(NOMA) and sparse code multiple access (SCMA) which require lesser signaling overhead and thus are more Energy efficient compared to Orthogonal access schemes which require more signaling to assign orthogonal resources. Another disadvantage of OFDMA related to Energy efficiency is that it has a Gaussian amplitude distribution and hence needs high resolution ADC which means higher power consumption.

**Advanced Internode co-ordination:**

As 5G network deployment strategy is focused around densification the problem of interference assumes special importance. Since interference is directly related to energy efficiency, solutions such as interference co-ordination schemes and Co-ordinated Multipoint (CoMP) are being proposed to reduce interference between users by exchanging information between the schedulers. As shown in [22], CoMP can be exploited to increase energy efficiency by optimally distributing RRH's in a distributed RR-centralized BBU- architecture.

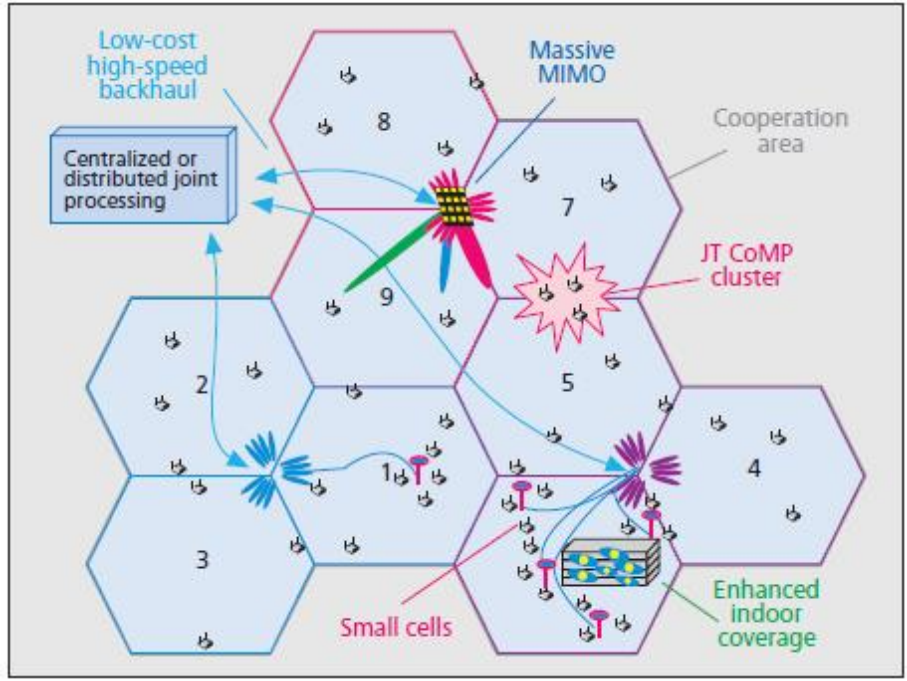

**Fig 6. Co-ordinated Multipoint type communications** *[Source: [1]]*

Relaying can help save energy in delay tolerant applications. For example a store-carry-and-forward (SCF) relay aided cellular architecture would mean that a device can transmit the data to a moving device which then retransmits it to the base station. It has been shown in [23] that a factor of more than 30 in savings can be obtained if such a scheme is used. Besides, relaying inherently saves energy due to the fact that it functions on short distance scales and so path losses are negligible. By operating at low power and generating less interference, they increase energy efficiency of the system.

Co-operative communications between devices is an extension of relay communication in that the devices relay information from the device to the BS but from BS to device too. One scenario where we can realize considerable savings is by exploiting the diversity of the relay channel. It is based on the assumption that while the link between device 1 and the BS may be bad, but overall link device 1-device 2-BS may not be bad. This diversity translates into energy savings as explained before. These energy savings can be increased if an appropriate modulation constellation size is chosen depending upon the distances we are targeting.

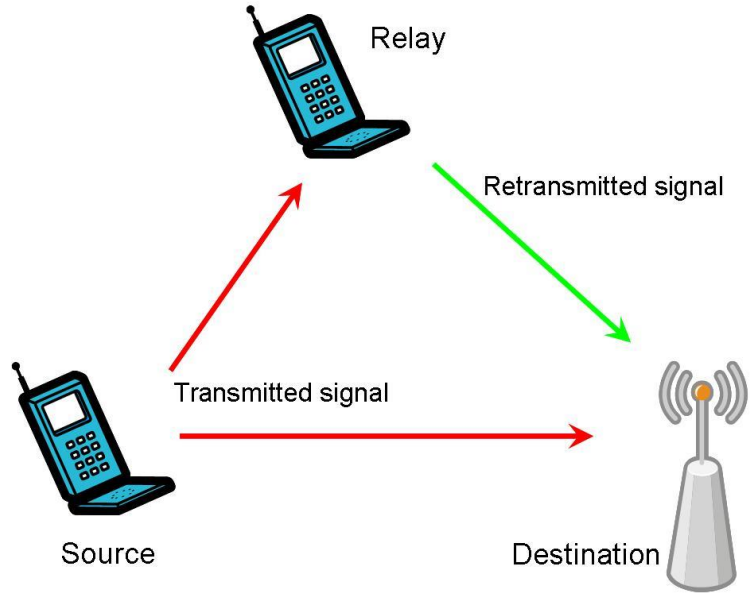

**Fig 7. Relay communications mechanism** *[Source: National Tsing Hua University]*

**Simultaneous Transmission and reception or Full duplex radio:**

All of the wireless systems today dedicate different spectral and time resources to Uplink (UL) and Downlink (DL) channels (Frequency Division Duplexing FDD or Time Division Duplexing TDD). Full duplex radio would enable bidirectional transmission and reception and thus save on spectrum. This by itself directly translates to energy savings. It would also play a major role in the control and signaling layers like conveying the CSI needed for SU-MIMO and MU-MIMO operation. This advantage also translates to energy savings.

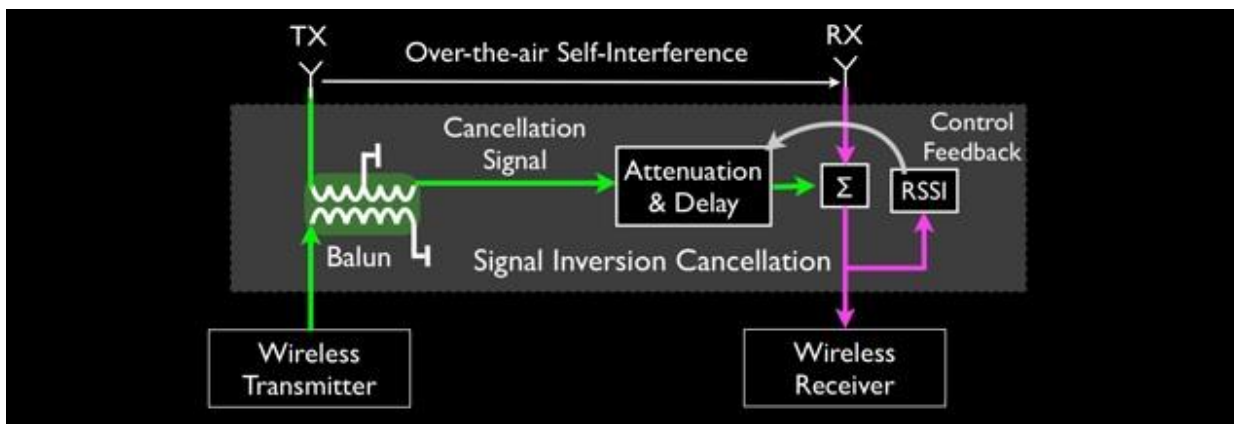

**Fig 7. Illustration of full duplex radio** *[Source: Stanford].*

**Het-nets and Multi-RAT's:**

With so much of legacy (2G, 3G, 4G) infrastructure around, it is clear that 5G will have to make good use of it. This would lead to heterogeneous networks and multi-rat integration and management would become critical. There would be dense deployments of small cells under it in an umbrella cell fashion. Further it has been proposed that the data plane and the control plane would be decoupled with the microcell handling the control and the small cell handling the data. This would allow the macrocell to turn off the small cell depending upon whether it is being used or not, leading to saving of power for that duration. A heterogeneous network would consist of 5G co-existing with all the legacy technologies 3GPP and non-3GPP and allow the user device to fall back on any legacy technology which could lessen its power drainage.

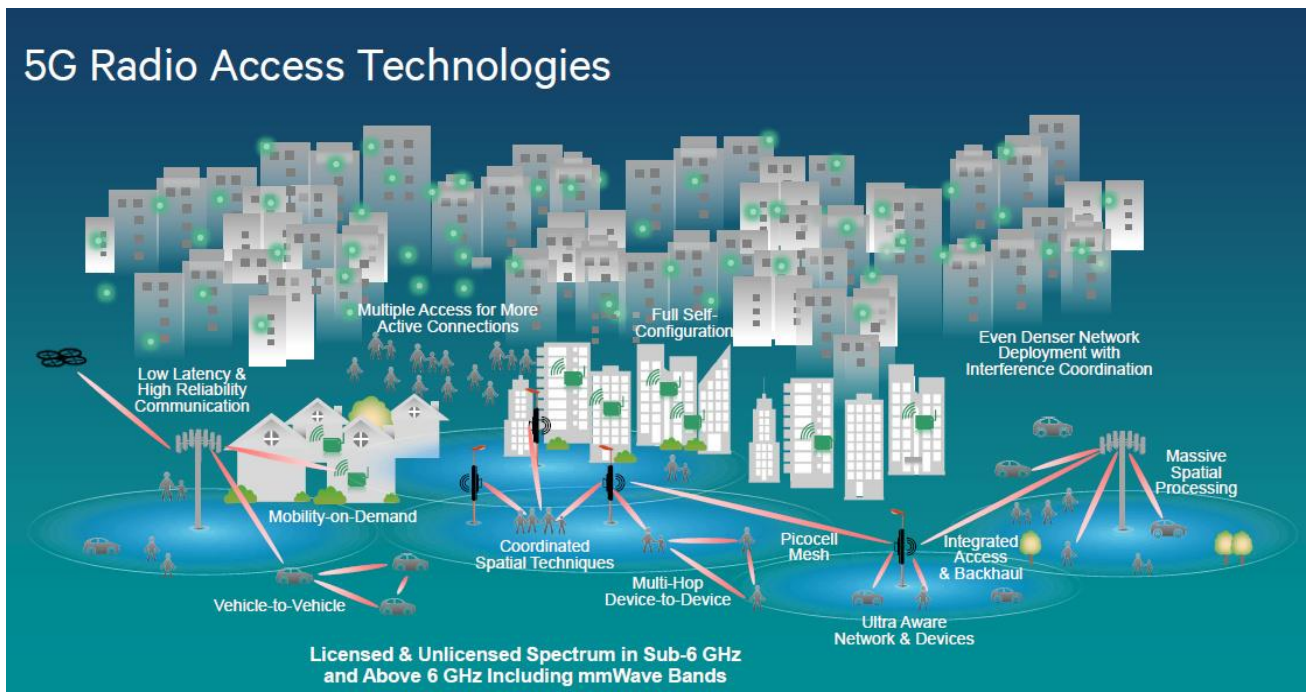

**Fig 8. A multi-RAT scenario** *[Source: Qualcomm]*

**Device-to-Device Communications:**

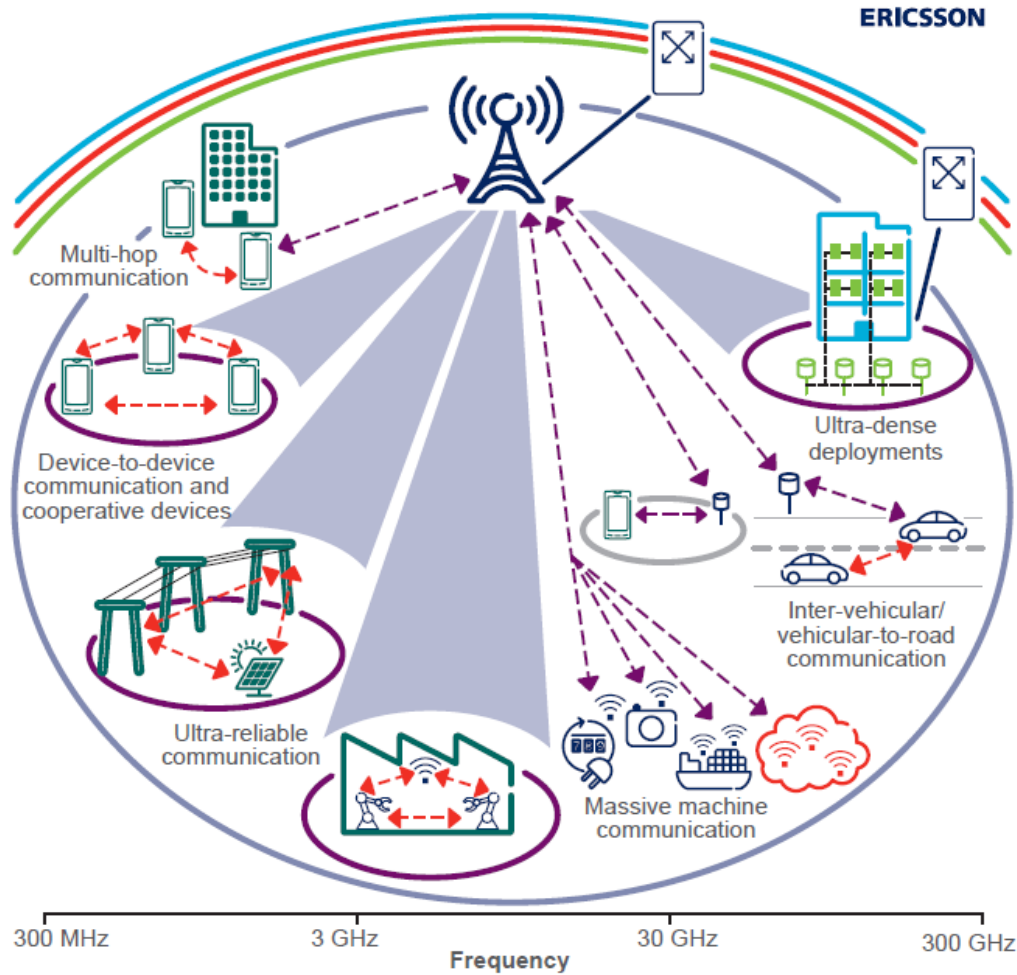

**Fig 9. Illustration of D2D and M2M type applications** *[Source: Ericsson]*

Device-to-Device (D2D) type communications that will be used for 2 reasons-

a. To extend coverage beyond the reach of access point, and

b. To reduce latency,

will operate in both licensed and unlicensed spectrum. Significant energy savings can be realized using this kind of an approach using technologies like CoMP, relay communications, co-operative communications discussed earlier, provided D2D is a well-integrated part of the overall access solutions [17].

**Machine-to-Machine Communications (M2M)/ Machine Type Communications (MTC) and IoT:**

Machine-to-Machine Communications (M2M) / Machine Type Communications (MTC) or IoT devices and applications and therefore communications will be characterized by large scale deployment, low bursty data rate requirement and low power. Therefore the 5G network will have to make special provisions for such services which could include reduced overhead for synchronization, channel allocation and connectivity management and putting the device into an idle state unless the device wants to transmit. Smartphone manufacturers use a similar feature where the smartphone makes or breaks connections with the network depending upon whether it wants to send data or not. This is known as fast dormancy. Such a DRX & DTX functionality in M2M and IoT communications will not only help save energy in the network but will extend the battery life of the device.

Thus the 5G network will have to be capable of supporting connectionless contention based access multiplexed with scheduled access on a common carrier with flexible allocation of resources dynamically. For eg. a UE in D2D mode, will need a scheduled capability but an M2M or IoT device would only need a connectionless capability to save power.

**Small Cells:**

Densification of 5G network will be achieved by the deployment of small low-power nodes as close as possible to the user. These base stations which will be deployed indoors and outdoors will help reduce the Capital expenditure (CAPEX) and Operating expenditure (OPEX) as they will operate on low power and the signals from such nodes will experience low path loss. Especially the ability to install them indoors, will avoid the need of wall penetration and consequently avoid power loss.

**Base stations:**

It is believed that energy consumption by base stations amounts to 70-80% of a cellular operator's expenditure on energy.

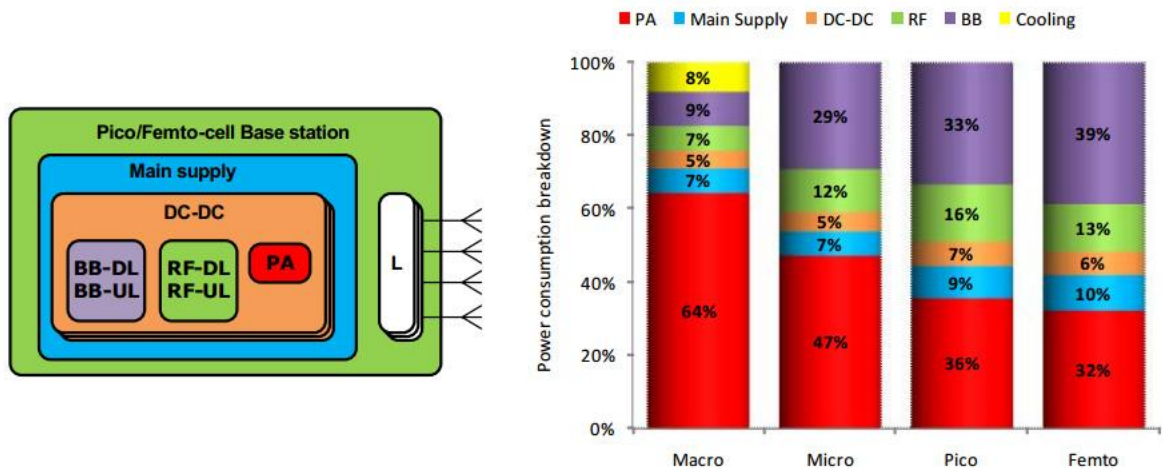

**Fig 10. A typical Pico/Femto cell base station with energy consumption breakdown of its various components.** *[Source: [12]].*

Fig 10. shows a typical base station transceiver. Each transceiver is made of BB interface, RF Transceiver, Power amplifier, a DC-DC power regulator, antennas and antenna interfaces. A base station has many such transceivers. Note that we are considering a base station with BBU and RRH collocated.

Fig 10. also shows the typical energy consumption in % of the different entities.

The changes in data traffic over the different periods of the day can cause under-utilization of the network. This idea can be exploited to save a significant amount of energy.

According to [12], power reduction of up to 20% can be obtained at low signal loads by adapting transmission parameters such as BW, modulation, coding rate, no.of antennas, duty cycle in time or frequency in the digital baseband engine.

Fig 11. shows the energy consumption of a 2*2 MIMO pico-cell BS BB Engine with traffic load with and without Energy Adaptation (EA).

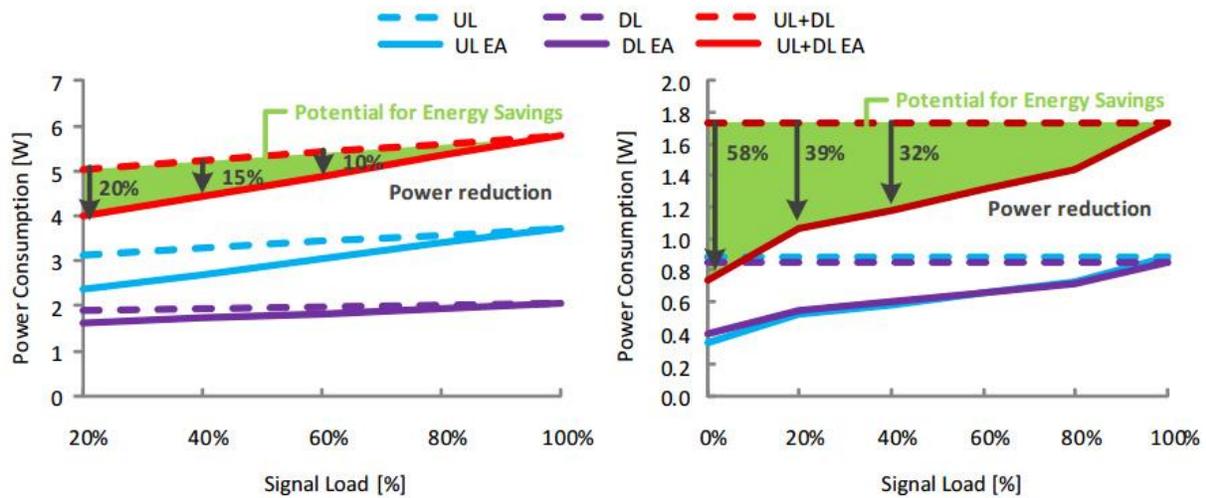

**Fig 11. Energy consumption of a 2*2 MIMO pico-cell BS BB Engine and RF transceiver with traffic load with and without Energy Adaptation (EA).** *[Source: [12]]*.

Also the RF transceiver can be made to adjust the SINAD performance to no better than what is required by the signal load. This can be done by implementing an analog Software Defined Radio (SDR) which can control the filtering (Bandwidth) and amplification (SINAD). To adapt according to signal load. Fig 11. also shows the energy consumption of a 2*2 MIMO pico-cell BS RF transceiver with traffic load with and without Energy Adaptation (EA).

Also according to [12], the savings can be realized in the RF power amplifier by adjusting the Operating point according to the required RF power level i.e. signal load and deactivation of PA stages when no power is required. Using a tunable matching network instead of a fixed matching network with the PA, increases PA efficiency and reduces power consumption. Similarly, a low loss RF end architecture such as the one shown in Fig 13. with SAW/BAW duplexers may be employed to further reduce power consumption.

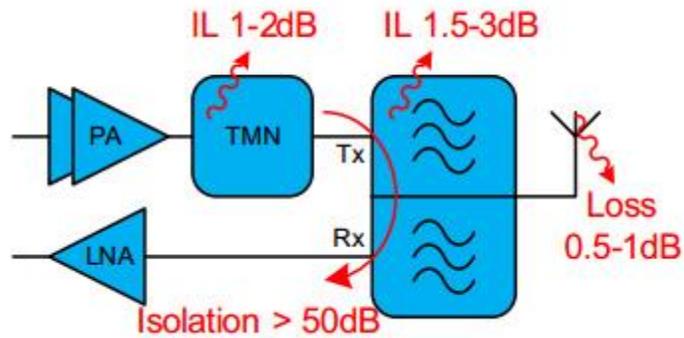

**Fig 12. Typical RF front-end architecture** *[Source [12]].*

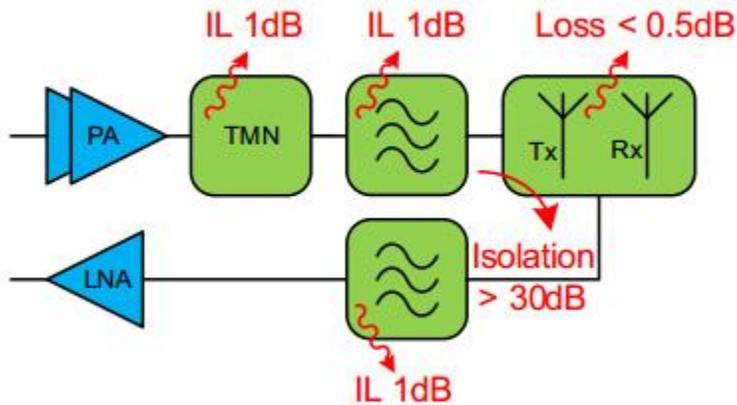

**Fig 13. Low loss RF front-end architecture with separate TX/RX antenna ports and relaxed filter specifications** *[Source [12]].*

**Backhauling:**

The importance of a green backhauling solution is a concern because studies have shown that the power saved by employing many low power base stations, ultimately contributes little to the savings because of the large amount power needed to backhaul the traffic from each one of the low power base stations [24]. Further the industry is divided on the choice of backhaul media. While some claim that millimeter waves used for access should also be used for backhaul while others feel fiber is a better choice for the backhaul.

However no study has been carried out till date to compare the performance of the 2 backhaul scenarios from an energy efficiency perspective.

**Context-Aware Networking:**

Context-Aware networking means that the network provide all users, devices, and applications with relevant services by taking into account many parameters like:

1. **Device context:** like device type, battery level, CPU load
2. **User context:** User preferences, User location, User history
3. **Application context:** real-time video streaming, video-on-demand, interactive gaming, web browsing
4. **Environmental context:** motion, lighting conditions, devices in vicinity
5. **Network context:** congestion in network, air-link quality, backhaul quality, spectrum availability etc. This kind of context awareness allows the device and the entire system to be more energy efficient in its functioning.

For e.g. m2M or IoT devices would not be provided mobility management and paging. The context aware network would know that for such a device low latency communication would suffice.

Inside of UE devices, as discussed in [20] an energy efficient network discovery algorithm may be used which decides to perform network scanning to discover new networks (rather than performing it periodically which is inefficient) and eventually vertically handing off to a lesser energy consuming network.

**Cross layer optimization:**

While the traditional layer-wise approach of independent Energy Efficient design of the 5G system can lead to higher design margins, a cross layer optimization approach would definitely do better in terms energy efficiency as well as present comparatively relaxed design margins since it exploits the interactions between the different layers of the protocol stack. Such a system-based approach is all the more relevant in a wireless system because of the more pronounced interactions between different entities due to the fact that the wireless medium is shared. Design of link layer, network layer, application layer using this approach has been studied in great detail in [26]. [25] discusses in great detail the cross layer design strategies in time, frequency and spatial domain.

**Network Function virtualization/Software defined networking/cloud/Flexible networks:**

The use of NFV, SDN and cloud is going to change telecommunications and networking altogether and 5G will be no exception to this. In fact these technologies, along with redistributed intelligence will be the core of 5G.

NFV is already being used widely in 4G. NFV involves executing functions which were traditionally executed on hardware, now on hypervisors in datacenters. With it, basically, NFV brings decoupling of the data and the control plane i.e. decoupling the services from the physical layer.

SDN on the other hand, provides an overlay for the virtualized functions. For e.g.: It can be used to control allocation of resources, implementing policies, QoS management, mobility management, security, charging etc. [17].

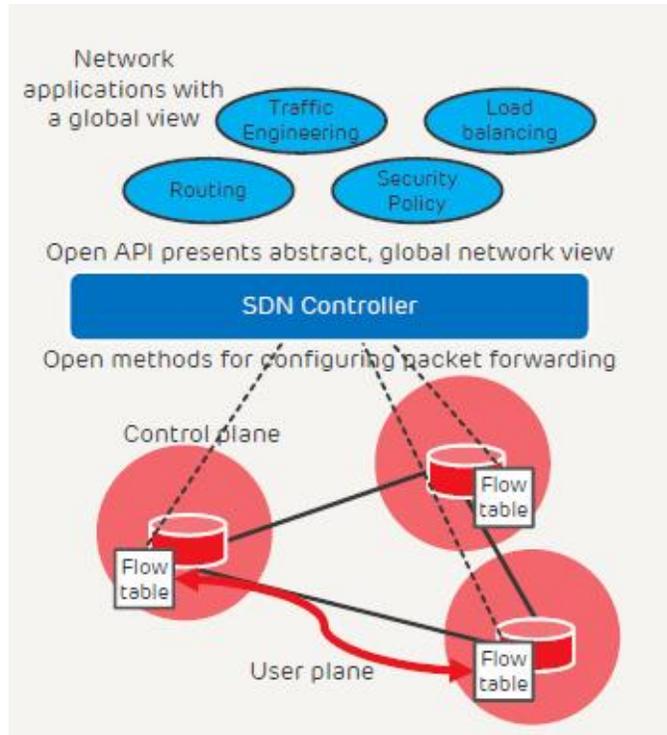

**Fig 14. Basic concept of SDN** *[Source: EE]*

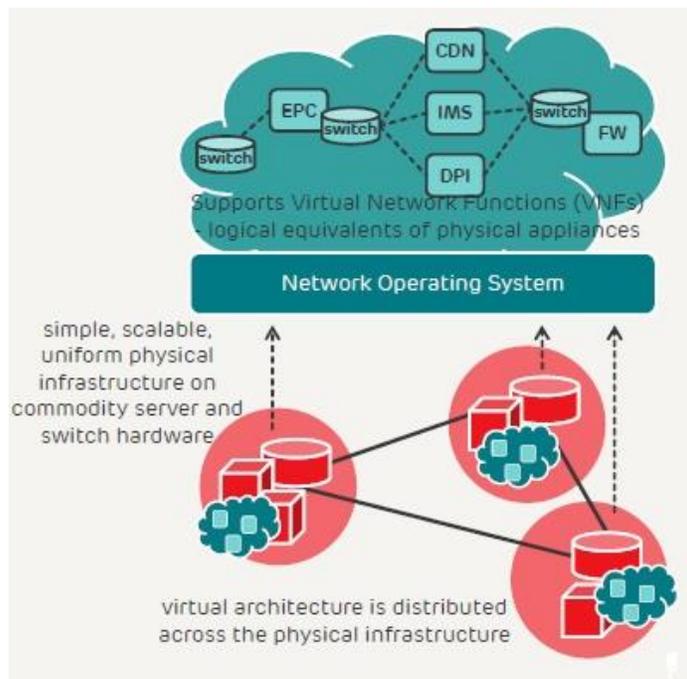

**Fig 15. Basic concept of NFV** *[Source: EE]*

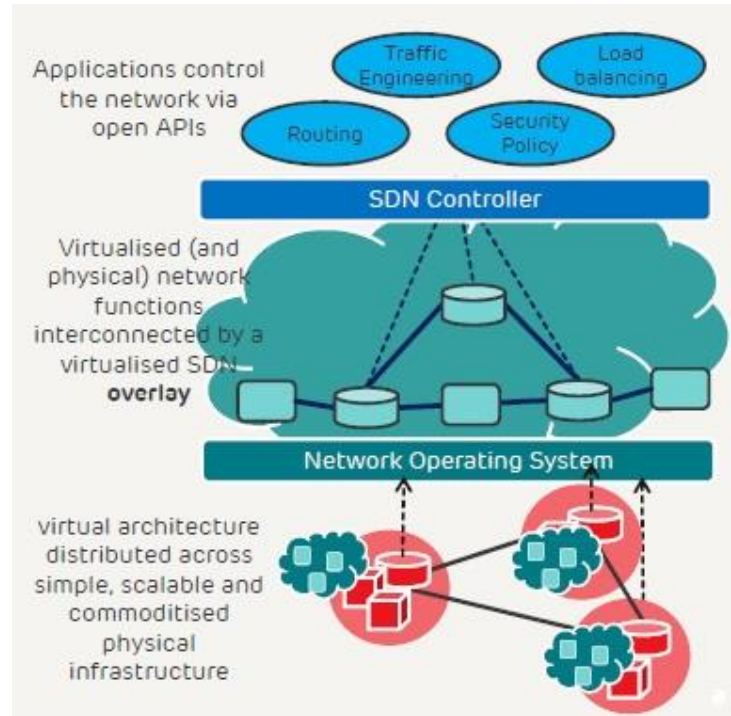

**Fig 16. Interoperation of SDN and NFV** *[Source: EE]*

Along with SDN and NFV, the use of orchestration technology to automate the deployment, co-ordination and management of network functions and services would provide multiple benefits to 5G. Energy efficiency is one of the most important of them.

Lesser hardware deployment and moving more and more functions to software would mean huge energy savings. Pushing these functions to datacenters would mean a lot of flexibility and energy savings. It is estimated that cooling accounts for about 50% of a datacenter's power usage. Common cooling systems would definitely help reduce it. Since the computing would be more centralized, energy saving techniques like power aware computing which could employ clock gating i.e. reducing clock speeds or power gating i.e. powering off parts of chips when idle, could be used. Scheduling and load balancing between servers or even shutting down a server etc. depending upon the computational load can bring significant energy changes. Centralization will also help attain a lower convergence time and a better global optimum.

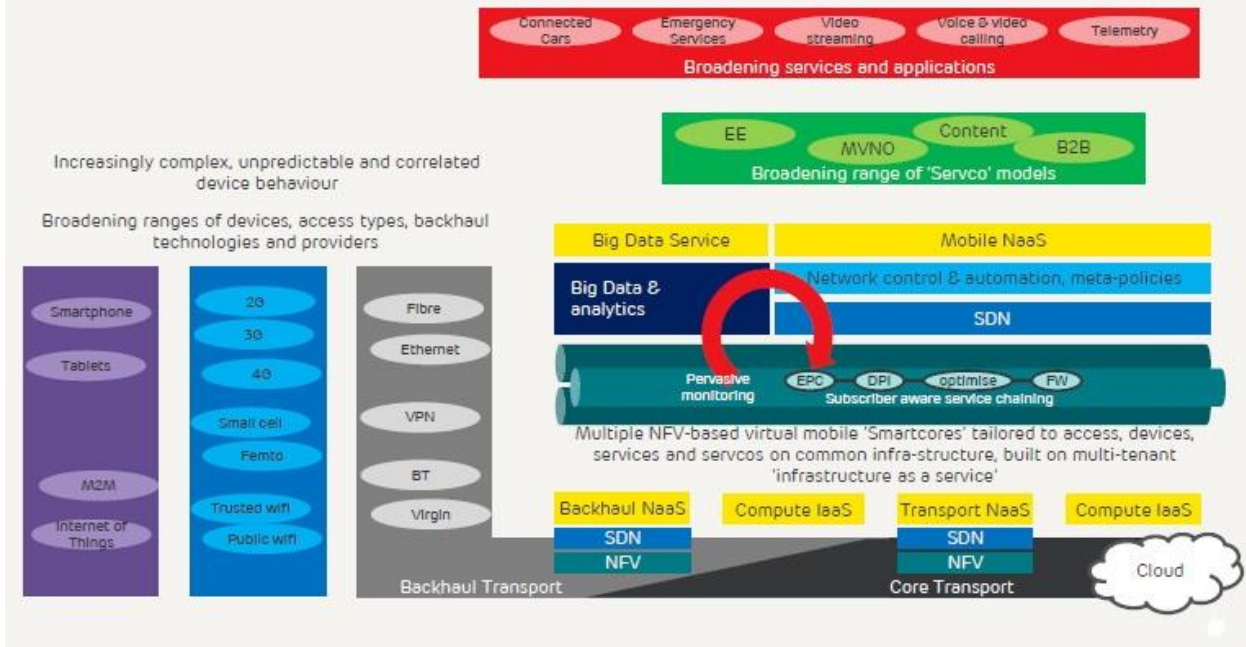

**Fig 17. How NFV and SDN will shape the future of cellular networks** *[Source: EE]*

All these NFV/SDN capabilities are in fact the technologies which will enable agile networks, C-RAN, switching on/off of RAN or backhaul as needed or for e.g. Multi-RAT selection based on a fuzzy logic based system taking UE battery status into account. All of these point to an increased energy efficiency. Since the network would be more software defined and have less hardware deployment, there would be lesser hardware failures in case of calamities for e.g. earthquakes, which would not bring down the network.

# CONCLUSION

Looking at the possible use cases of 5G, it is clear that 5G will touch every industry. Considering the energy problems, we are facing today it is required that our actions at least do not aggravate the problem further. It is therefore very important, that a technology like 5G which will be ubiquitous in the future, should be energy efficient. We have seen how the possible technology enablers of 5G already have a tinge of energy efficiency. But this is not enough. Our devices too, will have to operate on very, very low powers. Every 2 years, the processor power consumption is increasing 150% due to the application requirements detailed at the start of the document. But the battery capacity is increasing at a mere 10% every 2 years [4]. This is just one example out of the many. It is therefore clear that it would require multidisciplinary research in energy efficiency to ensure that 5G, when it comes, will be palatable.

# REFERENCES

[1] Jungnickel, V.; Manolakis, K.; Zirwas, W.; Panzner, B.; Braun, V.; Lossow, M.; Sternad, M.; Apelfröjd, R.; Svensson, T., "The role of small cells, coordinated multipoint, and massive MIMO in 5G,"*Communications Magazine, IEEE* , vol.52, no.5, pp.44,51, May 2014

[2] Li-Chun Wang; Rangapillai, S., "A survey on green 5G cellular networks," *Signal Processing and Communications (SPCOM), 2012 International Conference on* , vol., no., pp.1,5, 22-25 July 2012

[3] Olsson, M.; Cavdar, C.; Frenger, P.; Tombaz, S.; Sabella, D.; Jantti, R., "5GrEEn: Towards Green 5G mobile networks," *Wireless and Mobile Computing, Networking and Communications (WiMob), 2013 IEEE 9th International Conference on* , vol., no., pp.212,216, 7-9 Oct. 2013

[4] Daquan Feng; Chenzi Jiang; Gubong Lim; Cimini, L.J., Jr.; Gang Feng; Li, G.Y., "A survey of energy-efficient wireless communications," *Communications Surveys & Tutorials, IEEE* , vol.15, no.1, pp.167,178, First Quarter 2013

[5] Cavalcante, R.L.G.; Stanczak, S.; Schubert, M.; Eisenblaetter, A.; Tuerke, U., "Toward Energy-Efficient 5G Wireless Communications Technologies: Tools for decoupling the scaling of networks from the growth of operating power," *Signal Processing Magazine, IEEE* , vol.31, no.6, pp.24,34, Nov. 2014

[6] Chih-Lin I; Rowell, C.; Shuangfeng Han; Zhikun Xu; Gang Li; Zhengang Pan, "Toward green and soft: a 5G perspective," *Communications Magazine, IEEE* , vol.52, no.2, pp.66,73, February 2014

[7] Energy Efficient Networks Today and in the Future- Ericsson Research

[8] Boccardi, F; Heath, R.W.; Lozano, A.; Marzetta, T.L.; Popovski, P., "Five disruptive technology directions for 5G," *Communications Magazine, IEEE* , vol.52, no.2, pp.74,80, February 2014

[9] Bhushan, N.; Junyi Li; Malladi, D.; Gilmore, R.; Brenner, D.; Damnjanovic, A.; Sukhavasi, R.; Patel, C.; Geirhofer, S., "Network densification: the dominant theme for wireless evolution into 5G,"*Communications Magazine, IEEE* , vol.52, no.2, pp.82,89, February 2014

[10]   Wunder, G.; Jung, P.; Kasparick, M.; Wild, T.; Schaich, F.; Yejian Chen; Brink, S.; Gaspar, I.; Michailow, N.; Festag, A.; Mendes, L.; Cassiau, N.; Ktenas, D.; Dryjanski, M.; Pietrzyk, S.; Eged, B.; Vago, P.; Wiedmann, F., "5GNOW: non-orthogonal,

asynchronous waveforms for future mobile applications," *Communications Magazine, IEEE* , vol.52, no.2, pp.97,105, February 2014

[11]     Cheng-Xiang Wang; Haider, F.; Xiqi Gao; Xiao-Hu You; Yang Yang; Dongfeng Yuan; Aggoune, H.; Haas, H.; Fletcher, S.; Hepsaydir, E., "Cellular architecture and key technologies for 5G wireless communication networks," *Communications Magazine, IEEE* , vol.52, no.2, pp.122,130, February 2014

[12]     Debaillie, B.; Giry, A.; Gonzalez, M.J.; Dussopt, L.; Li, M.; Ferling, D.; Giannini, V., "Opportunities for energy savings in pico/femto-cell base-stations," *Future Network & Mobile Summit (FutureNetw), 2011* , vol., no., pp.1,8, 15-17 June 2011

[13]     Larsson, E.; Edfors, O.; Tufvesson, F.; Marzetta, T., "Massive MIMO for next generation wireless systems," *Communications Magazine, IEEE* , vol.52, no.2, pp.186,195, February 2014

[14]     Osseiran, A.; Boccardi, F.; Braun, V.; Kusume, K.; Marsch, P.; Maternia, M.; Queseth, O.; Schellmann, M.; Schotten, H.; Taoka, H.; Tullberg, H.; Uusitalo, M.A.; Timus, B.; Fallgren, M., "Scenarios for 5G mobile and wireless communications: the vision of the METIS project," *Communications Magazine, IEEE* , vol.52, no.5, pp.26,35, May 2014

[15]     Wooseok Nam; Dongwoon Bai; Jungwon Lee; Inyup Kang, "Advanced interference management for 5G cellular networks," *Communications Magazine, IEEE* , vol.52, no.5, pp.52,60, May 2014

[16]     Tehrani, M.N.; Uysal, M.; Yanikomeroglu, H., "Device-to-device communication in 5G cellular networks: challenges, solutions, and future directions," *Communications Magazine, IEEE* , vol.52, no.5, pp.86,92, May 2014

[17]     4G Americas' Recommendations on 5G Requirements and Solutions October 2014.

[18]     5G: A Technology Vision, Huawei.

[19]     5G and Wireless Broadband Evolution, Qualcomm.

[20]   Radwan Ayman, Rodriguez Jonathan, Energy Efficient Smart Phones for 5G Networks

[21]   Guowang Miao; Himayat, N.; Ye Li; Bormann, D., "Energy Efficient Design in Wireless OFDMA," *Communications, 2008. ICC '08. IEEE International Conference on* , vol., no., pp.3307,3312, 19-23 May 2008

[22]   Congqing Zhang; Tiankui Zhang; Zhimin Zeng; Cuthbert, L.; Lin Xiao, "Optimal Locations of Remote Radio Units in CoMP Systems for Energy Efficiency," *Vehicular Technology Conference Fall (VTC 2010-Fall), 2010 IEEE 72nd* , vol., no., pp.1,5, 6-9 Sept. 2010

[23]   Coll-Perales, B.; Gozalvez, J.; Friderikos, V., "Store, carry and forward for energy efficiency in multi-hop cellular networks with mobile relays," *Wireless Days (WD), 2013 IFIP* , vol., no., pp.1,6, 13-15 Nov. 2013

[24]   Verdú, S., "On channel capacity per unit cost," *Information Theory, IEEE Transactions on* , vol.36, no.5, pp.1019,1030, Sep 1990

[25]   G. Miao, N. Himayat, Y. Li, and A. Swami, "Cross-layer optimization for energy-efficient wireless communications: a survey," *Wireless Communications and Mobile Computing*, vol. 9, no. 4, pp. 529–542, 2009.

[26]   A. J. Goldsmith and S. B. Wicker, "Design challenges for energyconstrained ad hoc wireless networks," *IEEE Wireless Commun. Mag.*,

vol. 9, no. 4, pp. 8–27, 2002.